\documentclass[showpacs,aps,prd]{revtex4}
\input epsf

\begin{document}

\title{Study of $X_{c}(3250)$ as a $D_{0}^{*}(2400)N$ molecular state}
\author{Jian-Rong Zhang}
\affiliation{Department of Physics, College of Science, National University of Defense Technology,
Changsha 410073, Hunan, People's Republic of China}

%%%%%%%%%%%%%%%%%%%%%%%%%%%%%%%%%%%%%%%%%%%%%%%%%%%%%%%%%%%%%%%%%%%%%
\begin{abstract}
We present a QCD sum rule analysis for the
newly observed resonance $X_{c}(3250)$ by assuming it as
a $D_{0}^{*}(2400)N$ molecular state.
Technically, contributions of operators up to dimension $12$ are included in the
operator product expansion (OPE).
We find that it is difficult to find the conventional OPE convergence
in this work. By trying
releasing the rigid OPE convergence criterion,
one could find that the OPE convergence is still under control
in the present work and
the numerical result for $D_{0}^{*}(2400)N$ state is $3.18\pm0.51~\mbox{GeV}$,
which is in agreement with the experimental data of $X_{c}(3250)$.
In view of that the conventional OPE convergence
is not obtained here, thus only weak conclusions
can be drawn regarding the explanation of $X_{c}(3250)$
in terms of a $D_{0}^{*}(2400)N$ molecular state.
As a byproduct, the mass for the
bottom counterpart $\bar{B}_{0}^{*}N$ state is predicted to be $6.50\pm0.49~\mbox{GeV}$.
\end{abstract}
\maketitle

%%%%%%%%%%%%%%%%%%%%%%%%%%%%%%%%%%%%%%%%%%%%%%%%%%%%%%%%%%%%%%%%%%%%%
\section{Introduction}\label{sec1}
Very recently, BaBar Collaboration reported the measurement of the
baryonic $B$ decay $B^{-}\rightarrow\Sigma_{c}^{++}\bar{p}\pi^{-}\pi^{-}$
and observed a new
structure in the $\Sigma_{c}^{++}\pi^{-}\pi^{-}$ invariant mass spectrum at
$3.25~\mbox{GeV}$ \cite{X}.
For simplicity, one could name the new structure as $X_{c}(3250)$.
Soon after the experimental observation,
He {\it et al.} have suggested that $X_{c}(3250)$ could be
a $D_{0}^{*}(2400)N$ molecular state from
an effective Lagrangian calculation \cite{X-th}.
Theoretically, the molecular concept is well and truly not a new topic but with a history.
It was put forward nearly 40 years ago in Ref.
\cite{Voloshin} and was predicted that
molecular states have a rich spectroscopy in Ref. \cite{Glashow}.
The possible deuteron-like two-meson bound
states were studied in Ref. \cite{NAT}.
In recent years, some of ``X", ``Y", and ``Z" new
hadrons are ranked as possible molecular candidates.
Such as, $X(3872)$ could be a $D\bar{D}^{*}$ molecular
state \cite{X3872,X3872-1,X3872-2,X3872-3,X3872-4};
$X(4350)$ is interpreted as a $D_{s}^{*}\bar{D}_{s0}^{*}$ state
\cite{X4350-Zhang,X4350-Ma};
$Y(4260)$ is proposed to be
a $\chi_{c}\rho^{0}$ \cite{Y4260-Liu} or
an $\omega\chi_{c1}$  state \cite{Y4260-Yuan};
$Z^{+}(4430)$ is deciphered as a $D^{*}\bar{D}_{1}$ state
\cite{theory-Z4430n,theory-Z4430};
$Z_{b}(10610)$ and
$Z_{b}(10650)$ could be $B^{*}\bar{B}$ and $B^{*}\bar{B}^{*}$ states, respectively
\cite{Zb,Zb1}.
Especially, there already have a lot of works discussing baryon resonances
with meson-baryon molecular structures, e.g. \cite{new}.
If molecular states can be completely confirmed by experiment, QCD
will be further testified and then one will understand the QCD low-energy
behaviors more deeply.
Therefore, it is interesting to
study whether the newly observed $X_{c}(3250)$ state could be a $D_{0}^{*}(2400)N$ molecular state.

In the real world,
quarks are confined inside hadrons and the strong
interaction dynamics of hadronic systems
is governed by nonperturbative QCD effect completely.
Many questions concerning dynamics of the quarks and gluons at large distances remain
unanswered or understood only at a qualitative level.
It is quite difficult to extract hadronic information quantitatively
from the basic theory of
QCD. The QCD sum
rule method \cite{svzsum} is a nonperturbative
formulation firmly based on the first principle of QCD,
which has been successfully applied
to conventional hadronic systems, i.e. mesons or baryons
(for reviews
see \cite{overview1,overview2,overview3,overview4} and references
therein).
For multiquark states, there have appeared fruitful results from
QCD sum rules these years (for a review on multiquark QCD sum rules
one can see \cite{XYZ} and references
therein).
In particular for hadrons containing five quarks,
 some authors began to study
light pentaquark states in Refs. \cite{pentaquark}.
The application of QCD sum rules to heavy
pentaquark states was performed in Ref. \cite{pentaquark-heavy}
for the first time.

In this work, we devote to investigating that whether the newly observed resonance $X_{c}(3250)$ could be
a $D_{0}^{*}(2400)N$ molecular state ($D_{0}^{*}$ has a quark content $c\bar{q}$) in the framework of QCD sum rules.
As a byproduct, the mass for its
bottom counterpart $\bar{B}_{0}^{*}N$ is also predicted on the assumption that it could exist
($\bar{B}_{0}^{*}$ has a quark content $b\bar{q}$
and one should note that it has not been observed).
In theory,
one could expect that the $\bar{B}_{0}^{*}$ meson
should be assigned the same spin parity $0^{+}$ as $D_{0}^{*}$.
The rest of the paper is organized as three sections. We discuss QCD
sum rules for molecular states in Sec. \ref{sec2} utilizing similar techniques
as our previous works \cite{Zhang}.
The numerical analysis and discussions are presented in Sec. \ref{sec3},
and masses of $D_{0}^{*}(2400)N$ and $\bar{B}_{0}^{*}N$ molecular states are extracted
out. The Sec. \ref{sec4} includes
a brief summary and outlook.

%%%%%%%%%%%%%%%%%%%%%%%%%%%%%%%%%%%%%%%%%%%%%%%%%%%%%%%%%%%%%%%%%%%
\section{QCD sum rules for molecular states}\label{sec2}

The QCD sum rules for molecular states are constructed from the two-point correlation
function
\begin{eqnarray}\label{correlator}
\Pi(q^{2})=i\int
d^{4}x\mbox{e}^{iq.x}\langle0|T[j(x)\overline{j}(0)]|0\rangle.
\end{eqnarray}
In full theory, the interpolating current for $D_{0}^{*}(2400)$ or $\bar{B}_{0}^{*}$
meson can be found in Ref. \cite{reinders}, and
the one for nucleon $N$ has been listed in Ref. \cite{baryon-current}.
One can construct the $D_{0}^{*}(2400)N$ or $\bar{B}_{0}^{*}N$ molecular state current from
meson-baryon type of fields
\begin{eqnarray}
j&=&(\bar{q}^{c^{'}}Q^{c^{'}})(\varepsilon_{abc}q_{1}^{Ta}C\gamma_{\mu}q_{2}^{b}\gamma_{5}\gamma^{\mu}q_{3}^{c}),
\end{eqnarray}
where $Q$ is heavy quark $c$ or $b$, and $q$, $q_{1}$, $q_{2}$, as well as $q_{3}$ denote light quarks.
The index $T$ means matrix
transposition, $C$ is the charge conjugation matrix,
with $a$, $b$, $c$ and $c'$ are color indices.
One should note that meson-baryon molecules in the real world are long objects
in which the meson and the baryon are far away from each other.
The currents in this work and in most of the QCD sum rule works
are local and the five field operators here
act at the same space-time point. It is a limitation inherent
in the QCD sum rule disposal of the hadrons since the bound states are
not point particles in a rigorous manner.

Lorentz covariance implies that the
two-point correlation function in Eq. (\ref{correlator}) has the
form
\begin{eqnarray}
\Pi(q^{2})=\Pi_{1}(q^{2})+\rlap/q\Pi_{2}(q^{2}).
\end{eqnarray}
According to the philosophy of QCD sum rules, the correlator is
evaluated in two ways. Phenomenologically, the correlator can be
expressed as a dispersion integral over a physical spectral function
\begin{eqnarray}\label{pole-model}
\Pi(q^{2})=\lambda^{2}_H\frac{\rlap/q+M_{H}}{M_{H}^{2}-q^{2}}+\frac{1}{\pi}\int_{s_{0}}
^{\infty}ds\frac{\mbox{Im}\Pi_{1}^{\mbox{phen}}(s)+\rlap/q\mbox{Im}\Pi_{2}^{\mbox{phen}}(s)}{s-q^{2}}+\mbox{subtractions},
\end{eqnarray}
where $M_{H}$ is the mass of the hadronic resonance, and
$\lambda_{H}$ gives the coupling of the current to the hadron
$\langle0|j|H\rangle=\lambda_{H}u(p,s)$. In the OPE side, short-distance effects are taken care
of by Wilson coefficients, while long-distance confinement effects
are included as power corrections and parameterized in terms of
vacuum expectation values of local operators, the so-called
condensates.
One can write the correlation function in the OPE side
in terms of a dispersion relation
\begin{eqnarray}
\Pi(q^{2})=\int_{m_{Q}^{2}}^{\infty}ds\frac{\rho_{1}(s)}{s-q^{2}}+\Pi_{1}^{\mbox{cond}}(q^{2})+\rlap/q\bigg\{\int_{m_{Q}^{2}}^{\infty}ds\frac{\rho_{2}(s)}{s-q^{2}}+\Pi_{2}^{\mbox{cond}}(q^{2})\bigg\},
\end{eqnarray}
where the spectral density is given by the imaginary part of the
correlation function
\begin{eqnarray}
\rho_{i}(s)=\frac{1}{\pi}\mbox{Im}\Pi_{i}^{\mbox{OPE}}(s),~~i=1,2.
\end{eqnarray}
Technically, one works at leading order in $\alpha_{s}$ and considers condensates up
to dimension $12$. To keep the heavy-quark mass finite, one can use the
momentum-space expression for the heavy-quark propagator \cite{reinders}
\begin{eqnarray}
S_{Q}(p)&=&\frac{i}{\rlap/p-m_{Q}}
-\frac{i}{4}gt^{A}G^{A}_{\kappa\lambda}(0)\frac{1}{(p^{2}-m_{Q}^{2})^{2}}[\sigma_{\kappa\lambda}(\rlap/p+m_{Q})
+(\rlap/p+m_{Q})\sigma_{\kappa\lambda}]\nonumber\\&&{}
-\frac{i}{4}g^{2}t^{A}t^{B}G^{A}_{\alpha\beta}(0)G^{B}_{\mu\nu}(0)\frac{\rlap/p+m_{Q}}{(p^{2}-m_{Q}^{2})^{5}}[
\gamma^{\alpha}(\rlap/p+m_{Q})\gamma^{\beta}(\rlap/p+m_{Q})\gamma^{\mu}(\rlap/p+m_{Q})\gamma^{\nu}\\&&{}
+\gamma^{\alpha}(\rlap/p+m_{Q})\gamma^{\mu}(\rlap/p+m_{Q})\gamma^{\beta}(\rlap/p+m_{Q})\gamma^{\nu}+
\gamma^{\alpha}(\rlap/p+m_{Q})\gamma^{\mu}(\rlap/p+m_{Q})\gamma^{\nu}(\rlap/p+m_{Q})\gamma^{\beta}](\rlap/p+m_{Q})\nonumber\\&&{}
+\frac{i}{48}g^{3}f^{ABC}G^{A}_{\gamma\delta}G^{B}_{\delta\varepsilon}G^{C}_{\varepsilon\gamma}\frac{1}{(p^{2}-m_{Q}^{2})^{6}}(\rlap/p+m_{Q})
[\rlap/p(p^{2}-3m_{Q}^{2})+2m_{Q}(2p^{2}-m_{Q}^{2})](\rlap/p+m_{Q}).\nonumber
\end{eqnarray}
The light-quark part of the
correlation function can be calculated in the coordinate space, with the light-quark
propagator
\begin{eqnarray}
S_{ab}(x)&=&\frac{i\delta_{ab}}{2\pi^{2}x^{4}}\rlap/x-\frac{m_{q}\delta_{ab}}{4\pi^{2}x^{2}}-\frac{i}{32\pi^{2}x^{2}}t^{A}_{ab}gG^{A}_{\mu\nu}(\rlap/x\sigma^{\mu\nu}
+\sigma^{\mu\nu}\rlap/x)-\frac{\delta_{ab}}{12}\langle\bar{q}q\rangle+\frac{i\delta_{ab}}{48}m_{q}\langle\bar{q}q\rangle\rlap/x\nonumber\\&&{}\hspace{-0.3cm}
-\frac{x^{2}\delta_{ab}}{3\cdot2^{6}}\langle g\bar{q}\sigma\cdot Gq\rangle
+\frac{ix^{2}\delta_{ab}}{2^{7}\cdot3^{2}}m_{q}\langle g\bar{q}\sigma\cdot Gq\rangle\rlap/x-\frac{x^{4}\delta_{ab}}{2^{10}\cdot3^{3}}\langle\bar{q}q\rangle\langle g^{2}G^{2}\rangle,
\end{eqnarray}
which is then
Fourier-transformed to the momentum space in $D$ dimension.
The
resulting light-quark part is combined with the heavy-quark part
before it is dimensionally regularized at $D=4$.
Equating the two sides for $\Pi(q^{2})$ and assuming
quark-hadron duality yield the sum rules, from which masses of hadrons
 can be determined. After making a Borel transform and
transferring the continuum contribution to the OPE side, the sum
rules can be written as
\begin{eqnarray}\label{sumrule1}
\lambda^{2}_HM_{H}e^{-M_{H}^{2}/M^{2}}&=&\int_{m_{Q}^{2}}^{s_{0}}ds\rho_{1}(s)e^{-s/M^{2}}+\hat{B}\Pi_{1}^{\mbox{cond}},
\end{eqnarray}
\begin{eqnarray}\label{sumrule2}
\lambda^{2}_He^{-M_{H}^{2}/M^{2}}&=&\int_{m_{Q}^{2}}^{s_{0}}ds\rho_{2}(s)e^{-s/M^{2}}+\hat{B}\Pi_{2}^{\mbox{cond}},
\end{eqnarray}
where $M^2$ indicates the Borel parameter.
To eliminate the hadron coupling constant $\lambda_H$ and extract the
resonance mass $M_{H}$, one can take the derivative of Eq.
(\ref{sumrule1}) with respect to $1/M^{2}$, divide the result by itself
and deal with Eq. (\ref{sumrule2}) in the same way to get
\begin{eqnarray}\label{sum rule m}
M_{H}^{2}&=&\bigg\{\int_{m_{Q}^{2}}^{s_{0}}ds\rho_{1}(s)s
e^{-s/M^{2}}+d/d(-\frac{1}{M^{2}})\hat{B}\Pi_{1}^{\mbox{cond}}\bigg\}/
\bigg\{\int_{m_{Q}^{2}}^{s_{0}}ds\rho_{1}(s)e^{-s/M^{2}}
+\hat{B}\Pi_{1}^{\mbox{cond}}\bigg\},
\end{eqnarray}
\begin{eqnarray}\label{sum rule q}
M_{H}^{2}&=&\bigg\{\int_{m_{Q}^{2}}^{s_{0}}ds\rho_{2}(s)s
e^{-s/M^{2}}+d/d(-\frac{1}{M^{2}})\hat{B}\Pi_{2}^{\mbox{cond}}\bigg\}/
\bigg\{\int_{m_{Q}^{2}}^{s_{0}}ds\rho_{2}(s)e^{-s/M^{2}}
+\hat{B}\Pi_{2}^{\mbox{cond}}\bigg\},
\end{eqnarray}
where
\begin{eqnarray}
\rho_{i}(s)&=&\rho_{i}^{\mbox{pert}}(s)+\rho_{i}^{\langle\bar{q}q\rangle}(s)+\rho_{i}^{\langle\bar{q}q\rangle^{2}}(s)+
\rho_{i}^{\langle g\bar{q}\sigma\cdot G q\rangle}(s)+\rho_{i}^{\langle
g^{2}G^{2}\rangle}(s)+\rho_{i}^{\langle
g^{3}G^{3}\rangle}(s)+\rho_{i}^{\langle\bar{q}q\rangle^{3}}(s)+
\rho_{i}^{\langle\bar{q}q\rangle\langle
g\bar{q}\sigma\cdot G q\rangle}(s)\nonumber\\&&+
\rho_{i}^{\langle
g\bar{q}\sigma\cdot G q\rangle\langle
g\bar{q}\sigma\cdot G q\rangle}(s)+
\rho_{i}^{\langle\bar{q}q\rangle\langle
g^{2}G^{2}\rangle}(s)+
\rho_{i}^{\langle\bar{q}q\rangle\langle
g^{3}G^{3}\rangle}(s)+
\rho_{i}^{\langle
g^{2}G^{2}\rangle\langle
g\bar{q}\sigma\cdot G q\rangle}(s),~~i=1,2.
\end{eqnarray}
As a matter of fact, many terms of $\rho_{1}(s)$
are approximate to zero because they are
proportional to light quarks' masses in the calculations.
Thereby, we merely present the spectral densities
resulted from $\Pi_{2}(q^{2})$ here.
Concretely, they can be written as
\begin{eqnarray}
\rho_{2}^{\mbox{pert}}(s)&=&\frac{1}{3\cdot5^{2}\cdot2^{16}\pi^{8}}\int_{\Lambda}^{1}d\alpha \frac{(1-\alpha)^{6}}{\alpha^{5}}(\alpha s-m_{Q}^{2})^{4}(\alpha s+4m_{Q}^{2}),\nonumber\\
\rho_{2}^{\langle\bar{q}q\rangle}(s)&=&\frac{m_{Q}\langle\bar{q}q\rangle}{3\cdot2^{11}\pi^{6}}\int_{\Lambda}^{1}d\alpha\frac{(1-\alpha)^{4}}{\alpha^{3}}(\alpha s-m_{Q}^{2})^{3},\nonumber\\
\rho_{2}^{\langle\bar{q}q\rangle^{2}}(s)&=&\frac{\langle\bar{q}q\rangle^{2}}{3\cdot2^{8}\pi^{4}}\int_{\Lambda}^{1}d\alpha\frac{(1-\alpha)^{3}}{\alpha^{2}}(\alpha s-m_{Q}^{2})(\alpha s+m_{Q}^{2}),\nonumber\\
\rho_{2}^{\langle
g\bar{q}\sigma\cdot G q\rangle}(s)&=&-\frac{m_{Q}\langle
g\bar{q}\sigma\cdot G q\rangle}{2^{11}\pi^{6}}\int_{\Lambda}^{1}d\alpha\frac{(1-\alpha)^{3}}{\alpha^{2}}(\alpha s-m_{Q}^{2})^{2},\nonumber\\
\rho_{2}^{\langle
g^{2}G^{2}\rangle}(s)&=&\frac{m_{Q}^{2}\langle
g^{2}G^{2}\rangle}{5\cdot3^{2}\cdot2^{16}\pi^{8}}\int_{\Lambda}^{1}d\alpha\frac{(1-\alpha)^{6}}{\alpha^{5}}(\alpha s-m_{Q}^{2})(\alpha s-2m_{Q}^{2}),\nonumber\\
\rho_{2}^{\langle
g^{3}G^{3}\rangle}(s)&=&\frac{\langle
g^{3}G^{3}\rangle}{5\cdot3^{2}\cdot2^{18}\pi^{8}}\int_{\Lambda}^{1}d\alpha\frac{(1-\alpha)^{6}}{\alpha^{5}}[(\alpha s)^{2}-9\alpha sm_{Q}^{2}+10m_{Q}^{4}],\nonumber\\
\rho_{2}^{\langle\bar{q}q\rangle^{3}}(s)&=&\frac{m_{Q}\langle\bar{q}q\rangle^{3}}{3\cdot2^{4}\pi^{2}}\int_{\Lambda}^{1}d\alpha(1-\alpha),\nonumber\\
\rho_{2}^{\langle\bar{q}q\rangle\langle
g\bar{q}\sigma\cdot G q\rangle}(s)&=&-\frac{\langle\bar{q}q\rangle\langle
g\bar{q}\sigma\cdot G q\rangle}{2^{8}\pi^{4}}s\int_{\Lambda}^{1}d\alpha(1-\alpha)^{2},\nonumber\\
\rho_{2}^{\langle
g\bar{q}\sigma\cdot G q\rangle\langle
g\bar{q}\sigma\cdot G q\rangle}(s)&=&\frac{\langle
g\bar{q}\sigma\cdot G q\rangle^{2}}{2^{10}\pi^{4}}\int_{\Lambda}^{1}d\alpha(1-\alpha),\nonumber\\
\rho_{2}^{\langle\bar{q}q\rangle\langle
g^{2}G^{2}\rangle}(s)&=&\frac{m_{Q}\langle\bar{q}q\rangle\langle
g^{2}G^{2}\rangle}{3^{2}\cdot2^{13}\pi^{6}}\int_{\Lambda}^{1}d\alpha\frac{(1-\alpha)^{2}}{\alpha^{3}}[6\alpha^{2}(\alpha s-m_{Q}^{2})+(1-\alpha)^{2}(3\alpha s-4m_{Q}^{2})],\nonumber\\
\rho_{2}^{\langle\bar{q}q\rangle\langle
g^{3}G^{3}\rangle}(s)&=&-\frac{m_{Q}\langle\bar{q}q\rangle\langle
g^{3}G^{3}\rangle}{3\cdot2^{14}\pi^{6}}\int_{\Lambda}^{1}d\alpha\frac{(1-\alpha)^{4}}{\alpha^{3}},\nonumber\\
\rho_{2}^{\langle
g^{2}G^{2}\rangle\langle
g\bar{q}\sigma\cdot G q\rangle}(s)&=&-\frac{m_{Q}\langle
g^{2}G^{2}\rangle\langle
g\bar{q}\sigma\cdot G q\rangle}{3\cdot2^{13}\pi^{6}}\int_{\Lambda}^{1}d\alpha\frac{(1-\alpha)^{3}}{\alpha^{2}},\nonumber
\end{eqnarray}
and
\begin{eqnarray}
\hat{B}\Pi_{2}^{\mbox{cond}}&=&
-\frac{m_{Q}\langle\bar{q}q\rangle^{2}\langle
g\bar{q}\sigma\cdot G q\rangle}{2^{6}\pi^{2}}\int_{0}^{1}d\alpha e^{-m_{Q}^{2}/(\alpha M^{2})}
+\frac{m_{Q}^{2}\langle
g\bar{q}\sigma\cdot G q\rangle^{2}}{2^{10}\pi^{4}}\int_{0}^{1}d\alpha\frac{1-\alpha}{\alpha}e^{-m_{Q}^{2}/(\alpha M^{2})}\nonumber\\&&
+\frac{m_{Q}\langle\bar{q}q\rangle\langle
g^{2}G^{2}\rangle^{2}}{3^{3}\cdot2^{15}\pi^{6}}\int_{0}^{1}d\alpha\frac{(1-\alpha)^{2}}{\alpha}\bigg(\frac{3}{\alpha}-\frac{m_{Q}^{2}}{\alpha^{2}M^{2}}\bigg)e^{-m_{Q}^{2}/(\alpha M^{2})}\nonumber\\&&
+\frac{m_{Q}^{3}\langle\bar{q}q\rangle\langle
g^{3}G^{3}\rangle}{3^{2}\cdot2^{14}\pi^{6}}\int_{0}^{1}d\alpha\frac{(1-\alpha)^{4}}{\alpha^{4}}e^{-m_{Q}^{2}/(\alpha M^{2})}
+\frac{m_{Q}^{2}\langle\bar{q}q\rangle^{2}\langle
g^{2}G^{2}\rangle}{3^{3}\cdot2^{10}\pi^{4}}\int_{0}^{1}d\alpha\frac{(1-\alpha)^{3}}{\alpha^{2}}\bigg(\frac{2}{\alpha}-\frac{m_{Q}^{2}}{\alpha^{2}M^{2}}\bigg)e^{-m_{Q}^{2}/(\alpha M^{2})}\nonumber\\&&
+\frac{\langle\bar{q}q\rangle^{2}\langle
g^{3}G^{3}\rangle}{3^{3}\cdot2^{13}\pi^{4}}\int_{0}^{1}d\alpha\frac{(1-\alpha)^{3}}{\alpha^{2}}\bigg(\frac{11m_{Q}^{2}}{\alpha}-\frac{8m_{Q}^{4}}{\alpha^{2}M^{2}}\bigg)e^{-m_{Q}^{2}/(\alpha M^{2})}\nonumber\\&&
+\frac{\langle\bar{q}q\rangle\langle
g^{2}G^{2}\rangle\langle
g\bar{q}\sigma\cdot G q\rangle}{3^{2}\cdot2^{11}\pi^{4}}\int_{0}^{1}d\alpha\bigg[-1-\frac{(2-4\alpha+3\alpha^{2})m_{Q}^{2}}{\alpha^{3}M^{2}}+\frac{(1-\alpha)^{2}m_{Q}^{4}}{\alpha^{4}(M^{2})^{2}}\bigg]e^{-m_{Q}^{2}/(\alpha M^{2})}\nonumber\\&&
+\frac{m_{Q}^{3}\langle
g^{2}G^{2}\rangle\langle
g\bar{q}\sigma\cdot G q\rangle}{3^{2}\cdot2^{13}\pi^{6}}\int_{0}^{1}d\alpha\frac{(1-\alpha)^{3}}{\alpha^{3}}e^{-m_{Q}^{2}/(\alpha M^{2})}\nonumber\\&&
+\frac{m_{Q}\langle
g\bar{q}\sigma\cdot G q\rangle\langle
g^{3}G^{3}\rangle}{3^{2}\cdot2^{14}\pi^{6}}\int_{0}^{1}d\alpha\frac{(1-\alpha)^{3}}{\alpha^{2}}\bigg(\frac{3}{\alpha}-\frac{m_{Q}^{2}}{\alpha^{2}M^{2}}\bigg)e^{-m_{Q}^{2}/(\alpha M^{2})}
\end{eqnarray}
for $D_{0}^{*}(2400)N$ or $\bar{B}_{0}^{*}N$ state.
 The lower limit of  integration is
given by $\Lambda=m_{Q}^{2}/s$.

%%%%%%%%%%%%%%%%%%%%%%%%%%%%%%%%%%%%%%%%%%%%%%%%%%%%%%%%%%%%%%%%%%%
\section{Numerical analysis and discussions}\label{sec3}
In this section, the sum rule (\ref{sum rule q}) is numerically analyzed.
The input values are taken as
$m_{c}=1.23\pm0.05~\mbox{GeV}$, $m_{b}=4.24\pm0.06~\mbox{GeV}$,
$\langle\bar{q}q\rangle=-(0.23\pm0.03)^{3}~\mbox{GeV}^{3}$, $\langle
g\bar{q}\sigma\cdot G q\rangle=m_{0}^{2}~\langle\bar{q}q\rangle$,
$m_{0}^{2}=0.8\pm0.1~\mbox{GeV}^{2}$, $\langle
g^{2}G^{2}\rangle=0.88~\mbox{GeV}^{4}$, and $\langle
g^{3}G^{3}\rangle=0.045~\mbox{GeV}^{6}$
\cite{overview2}.

In order to ensure the quality of QCD sum rule analysis,
it is known that one can analyze the OPE convergence
and the pole contribution dominance to determine the conventional Borel window for $M^2$
in the standard QCD sum rule approach: on the one hand, the lower
constraint for $M^{2}$ is obtained by considering that the
perturbative contribution should be larger than each condensate
contribution to have a good
convergence in the
OPE side; on the other
hand, the upper bound for $M^{2}$ is obtained by the consideration
that the pole contribution should be larger than the continuum
state contributions. Meanwhile, the threshold
$\sqrt{s_{0}}$ is not arbitrary but characterizes the
beginning of continuum states.
Therefore, one naturally expects to find conventional Borel windows for studied states to make QCD sum rules work commendably.
However, things go contrary to one's wishes in some cases
and it may be difficult to find a conventional work window rigidly satisfying both of
two rules, which has been discussed in some works (e.g. Refs. \cite{Matheus,Zs}).
Referring to the present work, there also arises some similar problem. Concretely,
some condensates are very large and play an important role in the OPE side,
which makes the standard OPE convergence (i.e. the perturbative at least
larger than each condensate contribution) happen only at very large values of $M^2$.
The consequence is that it is difficult to find a conventional Borel window
where both the OPE converges well (the perturbative at least
larger than each condensate contribution) and the pole dominates over the continuum.

To obtain some useful hadronic information from QCD sum rules,
one could try
releasing the rigid convergence criterion of the perturbative contribution
larger than each condensate contribution in some case.
The comparison
between pole and continuum contributions from sum rule (\ref{sumrule2}) for $D_{0}^{*}(2400)N$ state
for $\sqrt{s_{0}}=3.8~\mbox{GeV}$ is shown in the left panel of FIG. 1, and its OPE convergence by comparing the
perturbative with other condensate contributions is shown in the right panel.
Not too bad for the present plight, there are four main condensates
(i.e. $\langle\bar{q}q\rangle$, $\langle
g\bar{q}\sigma\cdot G q\rangle$, $\langle\bar{q}q\rangle^{2}$,
and $\langle\bar{q}q\rangle\langle
g\bar{q}\sigma\cdot G q\rangle$)
and they could cancel out each other to some extent
since they have different signs. Besides,
most of other condensates calculated
are very small and almost negligible.
Thus, one could try
releasing the rigid OPE convergence criterion (i.e. the perturbative
larger than each condensate contribution) and restrict the ratio of the perturbative
to the ``total OPE contribution" (the sum of the perturbative and other condensates
calculated) at least larger than one half, for example $60\%$ or more. In other words,
here we consider the perturbative dominating over the sum of condensates
instead of the perturbative larger than each condensate.
Furthermore, it is also very important that we have examined that
condensates higher than dimension $12$
are quite small and the ratio
of the perturbative to the ``total OPE contribution" does not
change much even adding them (in
the total OPE contribution),
which means that condensates higher than dimension $12$
could not radically influence the character of OPE convergence here.
All the above factors
bring that the ratio of the perturbative
to the ``total OPE contribution" can be bigger than $60\%$ at relatively low values of $M^{2}$ in this work.
By way of parenthesis, one could also visually see that there exist very stable
plateaus from the Borel curves for the $D_{0}^{*}(2400)N$ state
shown in FIG. 2.

\begin{figure}[htb!]
\centerline{\epsfysize=5.8truecm\epsfbox{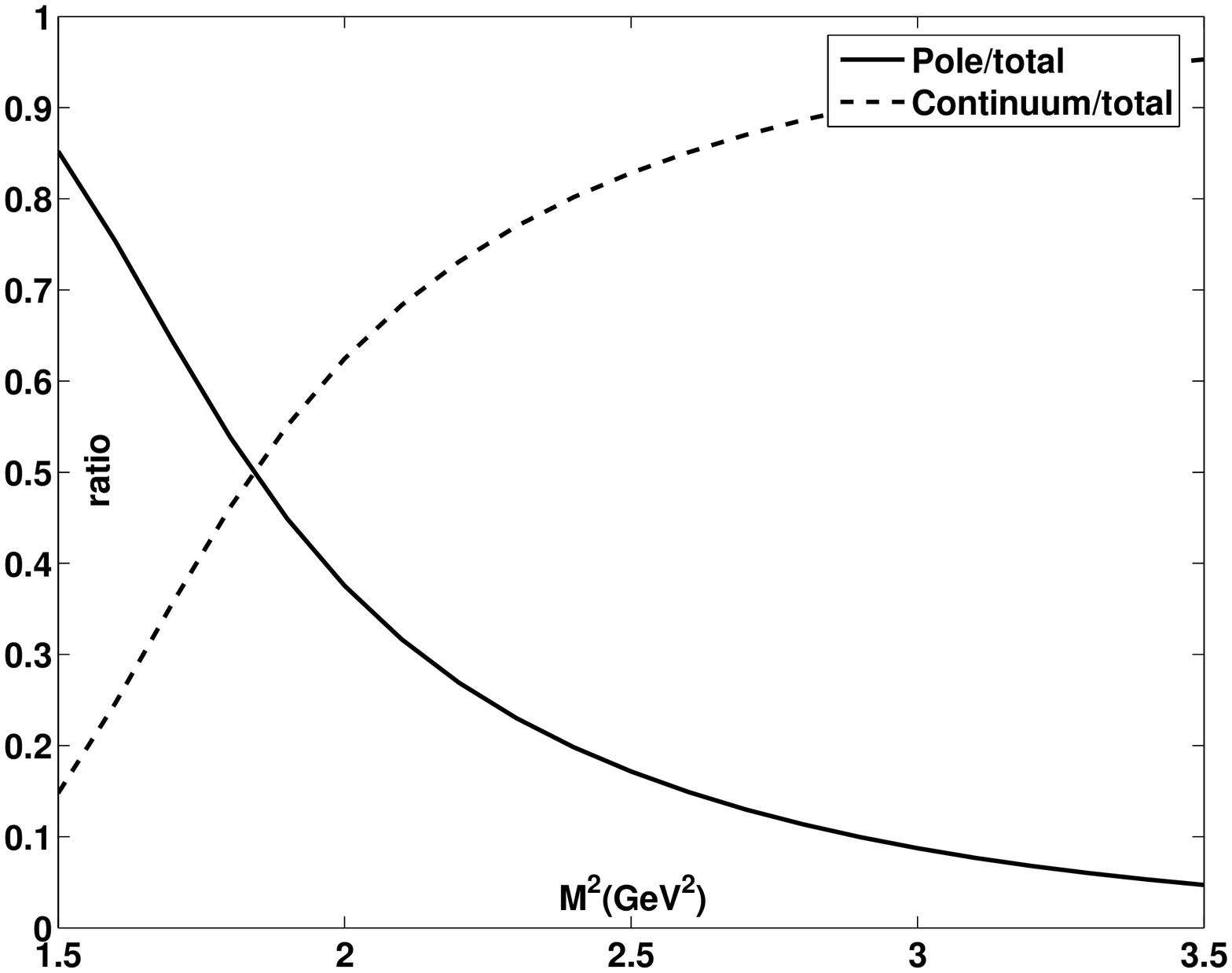}\epsfysize=5.8truecm\epsfbox{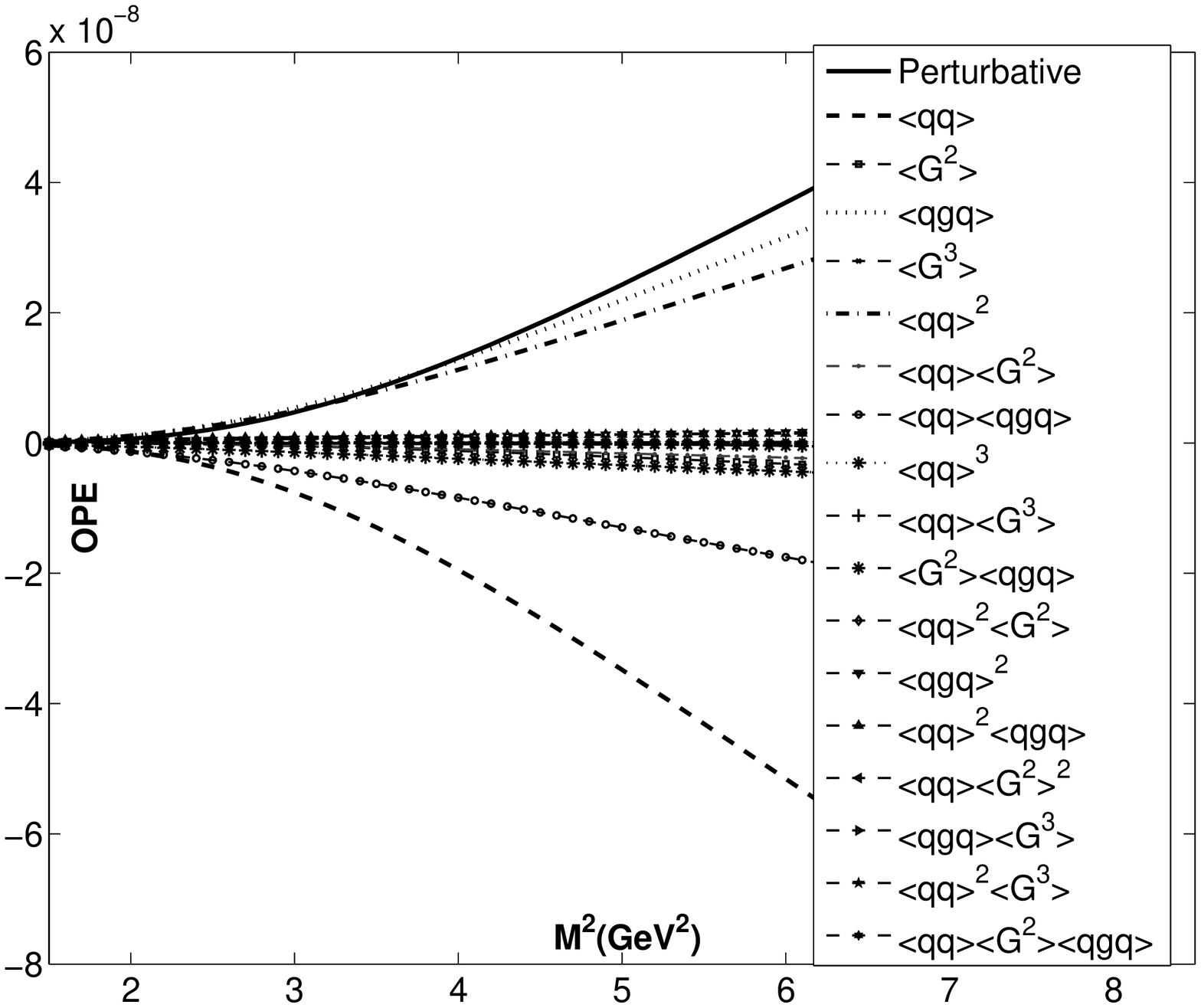}}
\caption{In the left panel, the solid line shows the relative pole contribution
(the pole contribution divided by the total, pole plus continuum
contribution) and the dashed line shows the relative continuum
contribution from sum rule (\ref{sumrule2}) for $\sqrt{s_{0}}=3.8~\mbox{GeV}$ for
$D_{0}^{*}(2400)N$ state. The OPE convergence is shown by comparing the
perturbative with other condensate contributions from sum rule (\ref{sumrule2}) for $\sqrt{s_{0}}=3.8~\mbox{GeV}$ for
$D_{0}^{*}(2400)N$ state in the right panel. }
\end{figure}

\begin{figure}
\centerline{\epsfysize=5.8truecm
\epsfbox{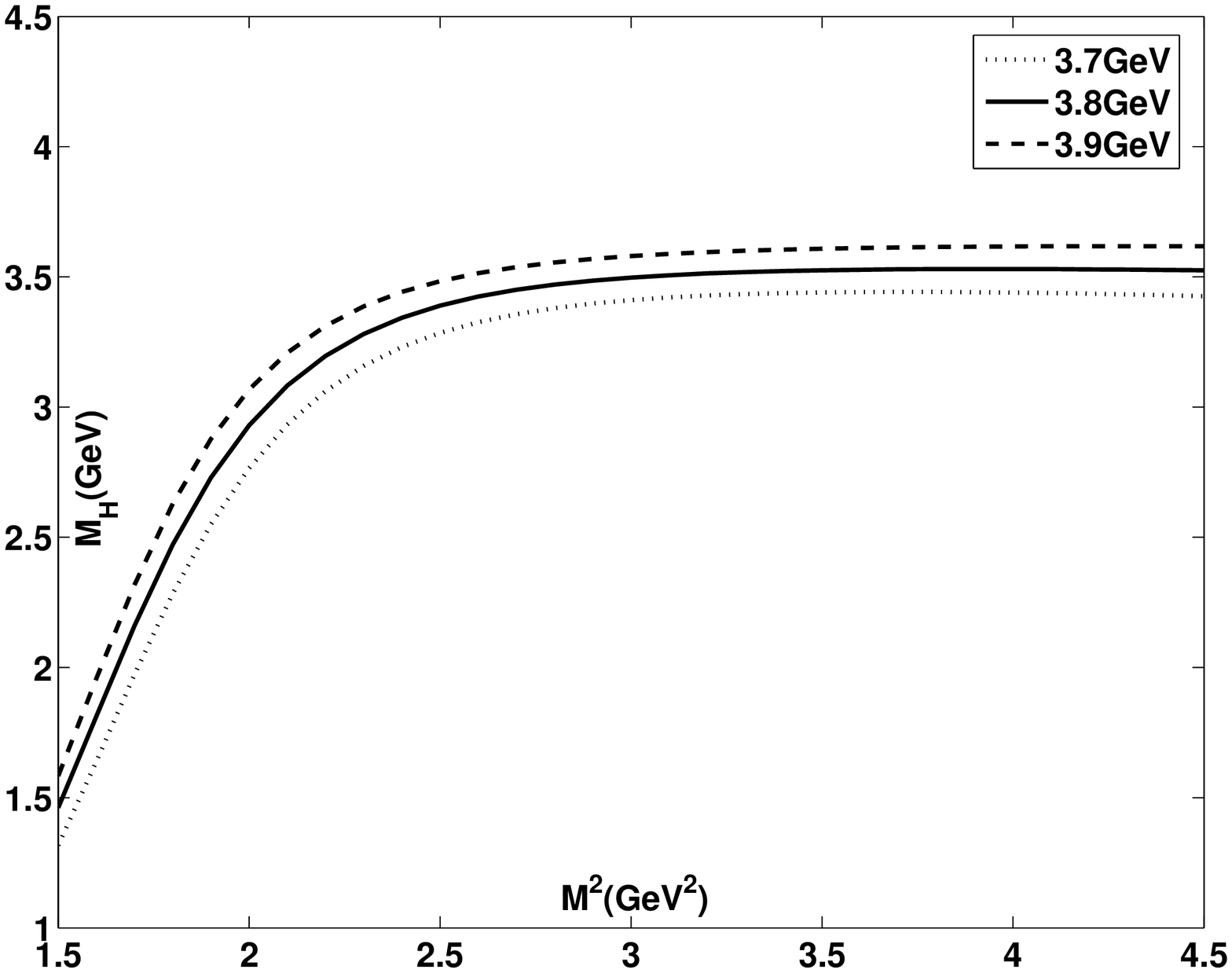}}\caption{The mass of $D_{0}^{*}(2400)N$ state as a function of $M^{2}$ from sum rule
(\ref{sum rule q}) is shown. The continuum
thresholds are taken as $\sqrt{s_0}=3.7\sim3.9~\mbox{GeV}$.}
\end{figure}

\begin{figure}
\centerline{\epsfysize=5.8truecm
\epsfbox{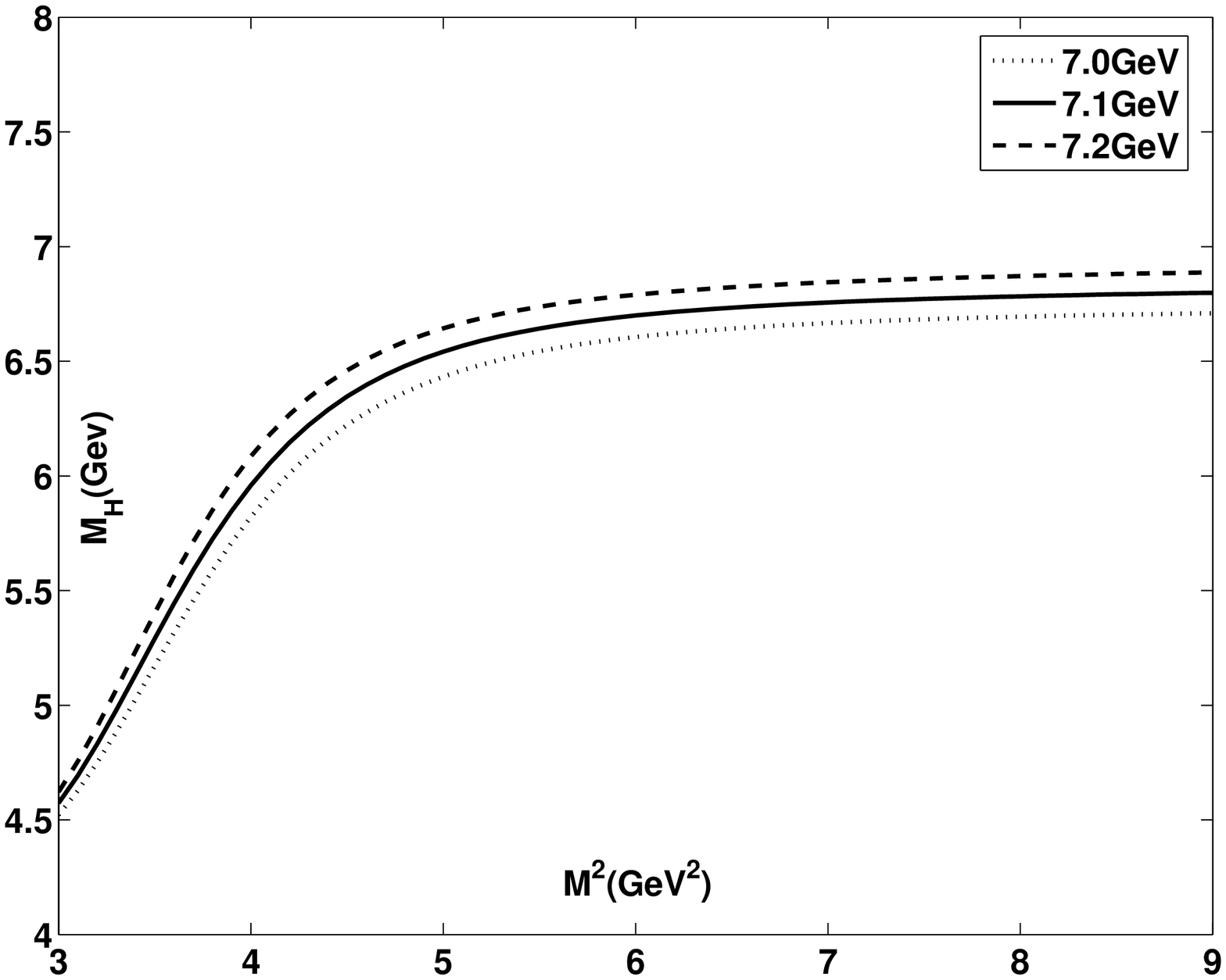}}\caption{The mass of $\bar{B}_{0}^{*}N$ state as a function of $M^{2}$ from sum rule
(\ref{sum rule q}) is shown. The continuum
thresholds are taken as $\sqrt{s_0}=7.0\sim7.2~\mbox{GeV}$.}
\end{figure}

All in all, to test the OPE convergence, we have considered the ratio of the perturbative
 to the ``total OPE contribution"
instead of the ratio of the perturbative to each condensate contribution.
This treatment is not freewheeling but has some definite constraints (i.e.
there are merely few important condensates
and they could cancel out each other to some extent; other condensates
are almost negligible).
In this sense, one could expect that
the OPE convergence is still under control.
We must truthfully admit that it is not a so good OPE convergence as the conventional
case, but then one could find a comparatively reasonable work window
and extract the hadronic information from QCD sum rules reliably.
Thus,
we choose some transition range $2.0\sim3.0~\mbox{GeV}^{2}$ as a compromise Borel window and
take
the continuum
thresholds as $\sqrt{s_0}=3.7\sim3.9~\mbox{GeV}$,
and arrive at
$3.18\pm0.41~\mbox{GeV}$ for $D_{0}^{*}(2400)N$ state.
Considering the uncertainty rooting in the variation of quark masses and
condensates, we gain
$3.18\pm0.41\pm0.10~\mbox{GeV}$ (the
first error reflects the uncertainty due to variation of $\sqrt{s_{0}}$
and $M^{2}$, and the second error resulted from the variation of
QCD parameters) or $3.18\pm0.51~\mbox{GeV}$
for $D_{0}^{*}(2400)N$ state.

On account of the difficulty encountered in finding a conventional Borel window,
one may suppose the nonexistence of $D^{*}_{0}(2400)N$ molecule itself.
As one possibility, the assumption of its nonexistence indeed should be drawn attention.
However, in the present work, we are inclined to make a premise that the $D^{*}_{0}(2400)N$ molecular state
could exist and then study whether it could act as one potential explanation of $X_{c}(3250)$ in view of two mian points:
I) The possibility for the existence of $D_{0}^{*}(2400)N$ molecule
and the molecular interpretation of $X_{c}(3250)$
are not entirely fabricated without any grounds.
By an effective Lagrangian calculation \cite{X-th}, He {\it et al.} found that $D_{0}^{*}(2400)$ and nucleon can form
a loosely bound state with the small binding energy, and
$X_{c}(3250)$ can be well explained as the $D_{0}^{*}(2400)N$ molecular hadron,
which is supported by
both the analysis of the mass spectrum and the study of its dominant
decay channel.
Moreover, the observed $X_{c}(3250)\rightarrow\Sigma_{c}^{++}\pi^{-}\pi^{-}$
can also be reasonably described.
II) We believe the present result from QCD sum rules
could provide another support to
the $D_{0}^{*}(2400)N$ explanation to $X_{c}(3250)$.
Certainly, we must confess to a weakness that it
is difficult to find the conventional Borel window in the present case.
Just as we have stated above, one could try
releasing the rigid OPE convergence criterion and eventually
find the OPE convergence is still under control
in the present case.
Although it is not a so good OPE convergence as the conventional
case, one could find a comparatively reasonable work window
and safely extract the hadronic information from QCD sum rules.

There comes forth the same problem for $\bar{B}_{0}^{*}N$ as the above case for $D_{0}^{*}(2400)N$,
and we treat it similarly.
The mass of $\bar{B}_{0}^{*}N$ state as a function of $M^{2}$ from sum rule
(\ref{sum rule q}) is
shown in FIG. 3.
Graphically, one can see there have very stable plateaus for Borel curves.
We choose a compromise Borel window $4.5\sim6.0~\mbox{GeV}^{2}$
and take $\sqrt{s_0}=7.0\sim7.2~\mbox{GeV}$
for $\bar{B}_{0}^{*}N$ state.
In the work windows, we obtain
$6.50\pm0.29~\mbox{GeV}$ for $\bar{B}_{0}^{*}N$ state. Varying input values of quark masses and
condensates, we attain
$6.50\pm0.29\pm0.20~\mbox{GeV}$ (the
first error reflects the uncertainty due to variation of $\sqrt{s_{0}}$
and $M^{2}$, and the second error resulted from the variation of
QCD parameters) or $6.50\pm0.49~\mbox{GeV}$
for $\bar{B}_{0}^{*}N$ state.

\section{Summary and outlook}\label{sec4}
Assuming the newly observed structure $X_{c}(3250)$ by BaBar Collaboration as
a $D_{0}^{*}(2400)N$ molecular state, we
calculate its mass value in the framework of QCD sum rules.
Technically, contributions of operators up to dimension $12$ are included in the
OPE.
We find that it is difficult to find the conventional OPE convergence
in this work. Via trying
releasing the rigid OPE convergence criterion, 
one could find that the OPE convergence is still under control
in the present work and
the final numerical result for $D_{0}^{*}(2400)N$ state is $3.18\pm0.51~\mbox{GeV}$,
which coincides with the experimental value $3.25~\mbox{GeV}$.
In view of that the conventional OPE convergence
is not obtained here, thus only weak conclusions
can be drawn regarding the explanation of $X_{c}(3250)$
in terms of a $D_{0}^{*}(2400)N$ molecular state.
Meanwhile, one should note that the $D^{*}_{0}(2400)N$ molecular state is just one
possible theoretical interpretation of $X_{c}(3250)$ and
there may have some other different explanations for its configuration.
One could expect that contributions from both future experimental observations
and theoretical analysis
will further reveal the nature structure of $X_{c}(3250)$.
Additionally, we have also studied the
bottom counterpart $\bar{B}_{0}^{*}N$ state
and predicted its mass to be $6.50\pm0.49~\mbox{GeV}$.
By analogy with $D_{0}^{*}(2400)N$ state,
this bottom counterpart state could be searched in the
$\Sigma_{b}\pi^{-}\pi^{-}$ invariant mass spectrum
in future experiments.
%%%%%%%%%%%%%%%%%%%%%%%%%%%%%%%%%%%%%%
\begin{acknowledgments}
This work was supported by the National Natural Science
Foundation of China under Contract Nos. 11105223, 10947016, 10975184, and the
Foundation of NUDT (No. JC11-02-12).
\end{acknowledgments}
%%%%%%%%%%%%%%%%%%%%%%%%%%%%%%%%%%%%%%%%%%%%%%%%%%%%

\end{document}